\documentclass[usenatbib]{mn2e}
\usepackage{amsmath}
\usepackage{graphicx}
\usepackage{textcomp}
\usepackage{txfonts}

\def\~{$\sim$}
\def\sn{\ensuremath{\bar{s}}}
\def\cacc{\ensuremath{\alpha_c}}
\def\wacc{\ensuremath{\alpha_w}}
\def\wpre{\ensuremath{\sigma_w}}
\def\wacccal{\ensuremath{\alpha_{w, calibrated}}}
\def\wprecal{\ensuremath{\sigma_{w, calibrated}}}
\def\wacczero{\ensuremath{\alpha_{w, 0}}}
\def\wprezero{\ensuremath{\sigma_{w, 0}}}
\def\diamean{\ensuremath{\mu_{\chi^2}}}
\def\diascat{\ensuremath{\sigma_{\chi^2}}}

\def\seeing{\ensuremath{\theta}}

\def\be{\begin{equation}}
\def\ee{\end{equation}}

\def\medlndiamean{\ensuremath{median\left(\ln{\diamean}\right)}}
\def\madlndiamean{\ensuremath{MAD\left(\ln{\diamean}\right)}}
\def\fbg{\ensuremath{F_{\rm bg}}}

\title[Reference image selection for difference imaging analysis]{Reference image selection for difference imaging analysis\thanks{The VVV test data used in this paper is based on observations made with European Southern Observatory Telescopes at the La Silla Paranal Observatory, under programme ID \mbox{179.B-2002}}}
\author[Huckvale, Kerins, Sale]{L. Huckvale,$^{1}$\thanks{E-mail:
leo.huckvale@postgrad.manchester.ac.uk} E. Kerins,$^{1}$ S. E. Sale$^{2}$
\\
$^{1}$Jodrell Bank Centre for Astrophysics, University of Manchester, M13 9PL, UK\\
$^{2}$Rudolf Peierls Centre for Theoretical Physics, University of Oxford, 1 Keble Road, Oxford, OX1 3NP, UK
}

\begin{document}

\date{Accepted 2014 April 25. Received 2014 April 24; in original form 2013 December 31}

\pagerange{\pageref{firstpage}--\pageref{lastpage}} \pubyear{2014}

\maketitle

\label{firstpage}

\begin{abstract}
Difference image analysis (DIA) is an effective technique for obtaining photometry in crowded fields, relative to a chosen reference image. As yet, however, optimal reference image selection is an unsolved problem. We examine how this selection depends on the combination of seeing, background and detector pixel size. Our tests use a combination of simulated data and quality indicators from DIA of well-sampled optical data and under-sampled near-infrared data from the OGLE and VVV surveys, respectively. We search for a figure-of-merit (FoM) which could be used to select reference images for each survey. While we do not find a universally applicable FoM, survey-specific measures indicate that the effect of spatial under-sampling may require a change in strategy from the standard DIA approach, even though seeing remains the primary criterion. We find that background is not an important criterion for reference selection, at least for the dynamic range in the images we test. For our analysis of VVV data in particular, we find that spatial under-sampling is best handled by reversing the standard DIA procedure and convolving target images to a better-sampled (poor seeing) reference image.
\end{abstract}


\section{Introduction}

Detection and time-resolved photometry of variable sources in globular clusters and the inner regions of galaxies is limited by stellar crowding. The most efficient method of detecting variables in these regions is difference imaging analysis (DIA, see \citealt{wozniak2008} for a review of the subject). In DIA, a reference image is chosen as a single image from a multi-epoch dataset, or built by stacking several images in that dataset. This is sometimes referred to as the ``template'' image in other literature. The images are geometrically aligned to the same coordinate system, resampling the data in each image to a common pixel grid. The reference image is then convolved by a kernel so that its point-spread function (PSF) matches that of each ``target'' image \citep{alard1998, alard2000}. Each target image is then subtracted from its corresponding convolved reference image to create a set of difference images. 

With each convolution (of the reference to a given target), one aims to minimise the residuals left in the difference image such that remaining residuals reflect true variable objects. For these objects, the residuals should correctly represent \emph{only} the difference in flux between the reference and the target epochs. Unwanted residuals may arise from spatial variations in the PSF across individual images due to changes in focus across the detector and, across wide fields, changes in seeing. Additionally, sky transparency or background illumination may vary across the images. Each part of the reference image must therefore be photometrically aligned to the corresponding part of the target image. Typically the convolution kernel is derived by fitting the coefficients of a set of basis functions by least-squares to minimise the difference image residuals. The kernel model used in the fit also includes terms for spatial variations in both the PSF and the background \citep{alard1998, alard2000}.

Once the difference images have been created, variable objects can be easily found by standard object detection on a stack of the absolute difference images. Photometry can then be carried out in each difference image in the time-series to build a difference-flux lightcurve for each object. DIA is particularly useful for obtaining high-precision photometry of objects for which brightness variations are a small fraction of average overall brightness \citep{alard1998}.

Selecting a good reference image for DIA is critical to obtaining accurate photometry. The reference image is common to every difference image created and is usually used to calibrate the final time-series photometry. The accuracy of photometry carried out on the difference images is therefore limited by the signal-to-noise in the reference image and the accuracy of the PSF-matching kernel convolution made prior to image subtraction. An accurate kernel solution requires sufficient information about the shape of the PSF in either the reference or target.

DIA has seen much development in the last few decades and has a generally agreed, mathematically robust algorithm, as will be outlined in \S\ref{sec:dia}. There is, however, \textit{no} such algorithm for selection of the reference image. A robust algorithm for reference image selection has potential for large gains in photometric accuracy. Automatic reference selection will also be useful for carrying out DIA with large-scale variability surveys, such as the VISTA Variables in the V\'ia L\'actea (VVV) survey \citep{minniti2010}, where manual reference selection is not desirable due to the quantity of data involved.

\citet{alard1998} advocate that the reference image should be the best-seeing image, as this has the highest signal to noise. However, this assumption has never been formally tested under varying regimes of pixelisation and dynamic range. Standard approaches to reference image selection do not account for images where the PSF is spatially under-sampled. Moreover, standard DIA methods (such as Optimal Image Subtraction; OIS, described in \S\ref{sec:dia-ois}) are not optimised for such data. Large-scale variability surveys may be subject to pixelisation, and thus under-sampling, due to survey strategy constraints on spatio-temporal coverage, as well as the nature of the instrumentation. In \S\ref{sec:dia-tic} we propose an alternative form of OIS in which target-images are convolved to match a poorer seeing reference image prior to subtraction, which could have some advantages when dealing with under-sampled data.

In \S\ref{calib} we assess the performance of different reference images under reference-convolving and target-convolving DIA, comparing the results of multiple DIA runs on two different datasets. Our test data, described in \S\ref{thedata}, are from the third-generation Optical Gravitational Lensing Experiment (OGLE-III; herein simply ``OGLE'') \citep{udalski2008} and VVV \citep{minniti2010}. In the OGLE test data, the PSF is spatially well-sampled, and the data have a high dynamic range between the typical background level and the saturation limit. The VVV test data chosen for this study frequently under-sample the PSF and have a low dynamic range. \S\ref{diaperf} describes the multiple DIA runs we performed to both compare reference-convolving and target-convolving DIA \emph{and} obtain DIA performance statistics against which we can test reference image selection algorithms.

We make two attempts to find a robust reference image selection algorithm. In \S\ref{sec:ufom}, we attempt to find a universal figure-of-merit (UFoM) which is applicable across all surveys and relates a set of image heuristics to expected DIA performance. With these heuristics, we intend to quantify the \emph{recoverability} of a PSF under varying conditions of pixelisation and signal-to-noise. They are calculated from simulations of a simple Gaussian PSF carried out over the space of simple image metrics which are easily obtained for potential reference images, namely seeing and background flux. In \S\ref{sec:sfom}, we attempt an empirical approach based on these simple image metrics alone. Here we test to what extent conventional wisdom (that a reference image should primarily be selected by seeing) applies across different regimes of well- or under-sampled data and, to some extent, high or low dynamic range (due to high background in the near-infrared and a low saturation limit).

\section{Difference imaging analysis}
\label{sec:dia}

\subsection{Optimal Image Subtraction} \label{sec:dia-ois}

The most widely used technique employed for DIA is the Optimal Image Subtraction (OIS) method, devised by \citet{alard1998} and further developed in \citet{alard2000}. We summarise this technique here.

The appropriate kernel, $K$, needed to convolve a region of the reference image, $R$, to match the seeing of the same region in the target image, $T$, is derived within a pixel grid centered on a bright star in that region. The difference image $D$ is formed through {\em reference image convolution}\/ (hereafter abbreviated as RIC):
\begin{equation}
\sum_i^{\mbox{\small reference}} D(x_i,y_i)^2 = \min \left[ \sum_{i} \left ( [R \otimes K](x_i,y_i) - T(x_i,y_i) \right )^{2} \right]
\label{eq:diffimage}
\end{equation}
over pixels $i$, where $(x_i, y_i)$ are the image pixel coordinates. 
This is the same as Equation 1 in \citet{alard2000}, except that we refer to the ``given'' image, $I$, as the ``target'' image, $T$. The width of the pixel grid should be chosen to be at least twice the seeing size (FWHM) so as to include most of the PSF.

The kernel may be modelled with a set of basis functions modified by two polynomials in the horizontal, $u$, and vertical, $v$, axes. The $u, v$ axes of the kernel are analogues of the $x, y$ axes of the real image. For the kernel model, \citet{alard1998} choose a set of two-dimensional circular Gaussian basis functions, $n$, multiplied by polynomials in $u$ and $v$ with orders $d_{n}^{x}$ and $d_{n}^{y}$, respectively. The kernel is thus the sum over all Gaussians $n$:
\begin{equation}
K(u,v)=\sum_{n} \sum_{d_{n}^{x}} \sum_{d_{n}^{y}} a_{n} \exp\left\{-\frac{(u^2+v^2)}{2\sigma_{n}^{2}}\right\} u^{d_{n}^{x}} v^{d_{n}^{y}}
\label{eq:kernel}
\end{equation}
One might choose the widths, $\sigma_{n}$, of each Gaussian to span some range about the estimated size of the PSF.

OIS is implemented in the ISIS software package, written by Christophe Alard\footnote{See: http://www.iap.fr/users/alard/package.html}. ISIS derives the spatial variation in the kernel by finding the local kernel within each of an array of small ``stamp'' regions distributed across the image. In each stamp, the local kernel is derived over a pixel grid centered on the brightest unsaturated star in the stamp. An overall kernel function for the image is then fitted to the coefficients of the basis functions, taking the coefficients themselves to be smoothly varying functions of $x$ and $y$. This is what makes OIS so powerful; fields which are \textit{more} crowded will have more bright stars across the image to inform the kernel derivation.

\subsection{Target image convolution} \label{sec:dia-tic}

Central to the optimal performance of OIS is the selection or construction of the reference image $R$ because this is imprinted onto every difference image.

\citet{tomaney1996} and \citet{alard1998} choose the highest signal-to-noise, best seeing images for their reference images. While \citet{tomaney1996} convolve all of their images (targets and reference) to a common, poor seeing image, \citet{alard1998} advocate convolving the reference image to match the seeing of each target image. This latter approach makes the best use of the signal-to-noise of the reference image, as well as that of each target image. It should be noted that \citet{tomaney1996} do not employ OIS, which came afterwards, and instead PSF-match their images via Fourier transforms.

Although OIS  takes account of spatially varying backgrounds, background flux must also be considered in making a choice of reference image, since a higher background flux in each pixel will have a greater contribution to the photon noise in that pixel, even after the background has been fitted and subtracted. Often it is desirable to stack several good seeing reference images to improve the signal-to-noise.

Maximising signal-to-noise by taking the best seeing image for the reference may seem the most intuitive option, but this approach relies on the PSF in the image being spatially well-sampled. The Nyquist sampling theorem maintains that to be able to reconstruct a continuous function from discrete data, one is required to sample that function at a frequency of at least twice the bandwidth occupied by its Fourier transform. Applying this to the spatial domain for a Gaussian PSF sampled by a pixel grid, one would require at least 2~pixels across the full-width half-maximum (FWHM) in order to reconstruct the PSF according to some Gaussian model function. Below this limit, the PSF recovered by fitting to the pixel values would be badly aliased. This is particularly a problem for image processing involving interpolation and resampling, such as geometric registration, carried out prior to difference imaging \citep{tomaney1996}. It is on this basis that \citet{wozniak2008} proposes a limit of 2.5~pixels~per~FWHM, below which images are ``under-sampled'' and could cause problems for PSF-matching. If the PSF transformation is not well determined then relative flux measurements will be subject to potentially significant systematic uncertainties.

Whilst DIA is now a well-developed and mature methodology, there is currently no optimal method for constructing the reference image nor for dealing with the special case of under-sampled data. The problems already described might suggest that one should abandon under-sampled data altogether when carrying out DIA, using only those images with seeing of more than 2.5~pixels. However, we are entering an age of wide-field time domain surveys which will require DIA to efficiently extract variability data from crowded fields. Survey time restrictions may force such surveys to operate with pixel scales near or under the sampling limit. Furthermore, near-infrared arrays typically have larger pixel scales than their optical counterparts as near-infrared array technology is less mature than CCD technology. It is therefore desirable to find a method of difference imaging and reference selection which can mitigate the problems of under-sampling.

One solution might be to flip the standard difference imaging equation~(\ref{eq:diffimage}) so that each target image is instead convolved to match the PSF of a worse-seeing reference image. The advantage of greater spatial sampling of the PSF in the reference image might outweigh the disadvantage of lower signal-to-noise during PSF-matching. Therefore, 
in our investigations of reference image selection, it is worth considering the effects of choosing a good or bad seeing reference image, either by performing RIC, as in equation~(\ref{eq:diffimage}), or {\em target image convolution}\/ (hereafter abbreviated as TIC):
\be 
   \sum_i^{\mbox{\small target}} D(x_i,y_i)^2 = \min \left[ \sum_{i} \left ( [T \otimes K](x_i,y_i) - R(x_i,y_i) \right )^{2} \right],
   \label{eq:targetdia}
\ee
where this time we convolve a target image to match a reference image with poorer seeing. 

There is an obvious disadvantage in constructing a difference image using TIC, as in equation~(\ref{eq:targetdia}), rather than RIC, as in equation~(\ref{eq:diffimage}). The main problem is the reference PSF in this case encircles more noise which is then imprinted onto all difference images. However this may be partially or fully mitigated by the advantages of this approach. By convolving to the same poor seeing image all difference images have the {\em same}\/ PSF (in the limit of a perfect kernel solution). This helps to make subsequent photometry simpler and reduces the effect of systematic errors. Additionally, most if not all of the resulting difference images can be directly stacked back onto the original reference image $R$ to make a new high signal-to-noise reference $R'$. Explicitly, for a fixed exposure time and given a sequence of $N$ target images $T_j$ convolved with kernels $K_j$ to the same PSF as $R$, the new reference $R'$ is obtained through
  \be
     R' =  \frac{1}{N+1} \left[ R + \sum_{j=1}^N (T\otimes K)_j \right] = R + \frac{1}{N+1} \sum_{j=1}^N D_j, \label{refstack}
  \ee
where $j$ denotes epoch. Equation~(\ref{refstack}) shows that $R'$ can be formed directly from $R$ and the difference image sequence $D_j$. In practise a more robust construction of $R'$ would be formed through inverse-variance weighting of $D_j$, although we omit this for the sake of clarity (i.e. we impose identical difference image variance). In the limit that the original kernel solutions $K_j$ converge towards their respective optimal solutions, the new set of difference images $D'_j$ derived from $R'$ can be obtained without a further convolution step since $R'$ is just a linear sum of existing difference images:
  \begin{eqnarray}
    D'_j = (T\otimes K')_j - R' & \simeq & (T\otimes K)_j - R' \nonumber \\
    & = & (D_j + R) - R' = D_j -  \frac{1}{N+1} \sum_{i=1}^N D_i, \label{conv}
  \end{eqnarray}
where we have omitted implied summations over pixels; indices $i$ and $j$ denote epoch and $K'$ is the new kernel solution derived from $R'$. In practise, one of the reasons for creating $R'$ will be to minimize noise in the reference and produce a new kernel solution $K'$ which will be superior to $K$. Also, we must bear in mind that for TIC DIA, equation~\ref{conv} implies that our new set of difference images would be formed from a stacked set of convolved images. This will likely lead to strongly correlated intra-pixel noise. However, equation~\ref{conv} at least presents a simple convergence test for obtaining the best possible kernel solution. Optimising strategies for stacked (as opposed to single-image) reference frames are beyond the scope of this paper and we do not consider them further here.

By contrast, RIC DIA involves convolving a good seeing reference image to the (poorer) PSF of each target image and so each difference image will have a different PSF. Whilst it is still generally advantageous to stack images to make a better signal-to-noise reference, typically only a small subset of the available images (those with superior seeing) can be used to do this.

Another potential major advantage of TIC DIA is the case where the PSF is under-sampled by the detector (as mentioned above), or where exposures may exceed the saturation level for bright stars under good seeing conditions. In fact both of these situation apply to VVV images as the seeing at Paranal is often around $0\farcs6$, compared to the VISTA camera pixel size of $0\farcs34$. Good difference imaging usually requires the PSF to be sampled by at least 2.5 pixels/FWHM. In crowded fields VVV $K_S$-band images also exhibit large numbers of saturated stars primarily due to the fairly low dynamic range available between the typical background level and the array saturation level. In this case it may well make better sense to use TIC in order to have both the most reliable kernel solutions and also to minimise the effects of saturation.

\section{Measuring reference image performance} \label{calib}

Ultimately, the quality of the final lightcurves provides the appropriate measure of how all stages of a photometry pipeline perform. However, lightcurve quality indicators are clearly sensitive to inefficiencies in any part of the pipeline, not just DIA. For this reason, our measure of reference image performance is based on indicators of image subtraction quality only. To this end, our analysis forms only part of the overall optimisation required to produce the best possible lightcurves.

In order to construct a figure-of-merit for reference image selection (\S\ref{sec:ufom}, \S\ref{sec:sfom}) we first need to determine the relative performance of reference images when difference imaging each against a range of target images, as a function of image quality variables. We therefore carried out multiple difference imaging runs with ISIS using test data described below. Each image in each test dataset was, in turn, used as a reference against the rest of the dataset. We carried out such tests using both RIC as given by equation~(\ref{eq:diffimage}) and TIC using equation~(\ref{eq:targetdia}). 

\subsection{The datasets} \label{thedata}

To test difference imaging performance we use 24 epochs of optical $I$-band data from the OGLE-III survey and 13 epochs of near-infrared $K_S$-band images from the VVV survey. In the OGLE images, stellar PSFs are well-sampled, with typical seeing of more than $3.5$~pixels. In the VVV data, seeing is typically less than $3$~pixels and stellar PSFs are often under-sampled, with seeing down to 1.3~pixels. Background levels in the near-IR VVV images are larger than those in the optical OGLE images and the saturation levels are lower giving a lower dynamic range for signals to be detected above the noise level. The two sets of images therefore present quite different characteristics for difference imaging.

\begin{table*}
\caption{Properties of the OGLE and VVV datasets used in this study.}
\label{tab:dataproperties}
\begin{tabular}{lcc}
\hline
						                        &	OGLE						                    	&	VVV							                            \\
\hline
Epochs						                    &	24 (March 2004 -- April 2009)	                    &	27 (May 2011 -- March 2013)			                    \\
Survey field designation			            &	BLG206							                    &	b292 		                                            \\
Field centre coordinates (galactic, J2000)	    &	\mbox{$l=1.6324$, $b=-2.6606$}	                    &	\mbox{$l=0.87$ $b=-3.23$}				                \\
Field centre coordinates (equatorial, J2000)	&	$\alpha=17^h59^m52.8$, $\delta=-28\degr53'00''$		&	$\alpha=18^h00^m25$, $\delta=-29\degr49'35.7''$		    \\
Photometric band				                &	$I$							                        &	$K_s$							                        \\
Camera						                    &	OGLE-III						                    &	VIRCAM							                        \\
Detectors					                    &	$2048\times4096$ pixel SITe CCD				        &	$2048\times2048$ pixel Raytheon VIRGO infrared array	\\
Detector number					                &	1 of 8							                    &	5 of 16							                        \\
Pixel scale (arcsec~pixel$^{-1}$)		        &	0.26							                    &	0.34							                        \\
Saturation limit (ADU)		                    &	65535							                    &	24000							                        \\
Image subsection used in this study (pixels)	&	$600\times600$						                &	$600\times600$						                    \\
\hline
\end{tabular}
\end{table*}

The OGLE dataset comprises a random selection of 24 sub-field images from the field ``BLG206'' obtained between March 2004 and April 2009, with seeing ranging from 3.50 to 6.49 pixels. The full $35'\times35'$ BLG206 field is centered on \mbox{$l=1.6324$, $b=-2.6606$} ($\alpha=17^h59^m52.8$, $\delta=-28\degr53'00''$ J2000). The images were obtained in the $I$ band with detector number 1 of the eight $2048\times4096$ pixel SITe CCDs which made up the OGLE-III camera, having a pixel scale of 0.26~arcsec~pixel$^{-1}$ and a saturation limit of 65535 ADU. The test data for this paper comprises a $600\times600$ pixel subsection cropped from the images after performing an approximate geometrical alignment.
The VVV dataset comprises 27 images from the field ``b292'' obtained between May 2011 and March 2013, with seeing ranging from 1.34--2.99 pixels. The full b292 field is centered on \mbox{$l=2.08$ $b=-3.08$} ($\alpha=18^h00^m25$, $\delta=-29\degr49'35.7''$ J2000). The images were obtained in the $K_s$ band with array 5 of the sixteen $2048\times2048$ pixel Raytheon VIRGO infrared arrays which make up the VIRCAM camera, having a pixel scale of 0.34~arcsec~pixel$^{-1}$ and a saturation limit (specific to array 5) of 24000 ADU. Again, the test data for this paper comprises a $600\times600$ pixel subsection of these images.
The properties of these datasets are summarised in Table \ref{tab:dataproperties}.

Seeing values for both datasets were taken from the FITS headers. The background flux per pixel was calculated from the modal pixel value for VVV and from the reported background in the FITS headers for OGLE, which is in excellent agreement with our modal pixel value (within $<1\%$), ranging from 627 to 1268 counts per pixel. For VVV data background levels are much larger, ranging from 3437 to 6011 counts per pixel due to the lessened effects of extinction in the $K_S$ band. The median flux of the brightest 1\% of stars in each image was 204000 for OGLE and 39000 for VVV, as summarised in Table~\ref{tab:metrics} (these values were determined by simple aperture photometry and rounded to the nearest $10^3$ ADU).

The differing saturation, background levels and seeing ranges between the OGLE and VVV datasets provides a good testbed for our analysis.

\begin{table}
\caption{Metrics derived for OGLE and VVV test datasets. The ranges in seeing, background and bright star flux span the variation in their value across all the images. The median bright star flux is used to compare with the PSF simulations in \S\ref{sec:psfsim}.}
\label{tab:metrics}
\begin{tabular}{lcc}
\hline
				&	OGLE	&	VVV		\\
\hline
Seeing (pixels)			&	3.50--6.49	&	1.68--2.99	\\
Background (ADU per pixel)	&	627--1268	&	6865--12042	\\
Median bright star flux (ADU per star)	&	204000		&	39000		\\
Saturation limit (ADU per pixel)	&	65535	&	24000\\
\hline
\end{tabular}
\end{table}

\subsection{Difference imaging performance} \label{diaperf}

When performing difference imaging, ISIS computes a reduced chi-square as an indicator of difference image quality, which for RIC is given by:
\begin{equation}
\chi^2_{red} = \frac{1}{N_{pixels}} \sum\limits_{i}^{N_{pixels}}{ \frac{\left ( (R \otimes k)_i - T_i + B \right )^2}{T_i} },\label{diaqual}
\end{equation}
where $N_{pixels}$ is the number of pixels within the stamp region over which the convolution kernel is evaluated and $B$ is the background term, which is taken to be constant over the region of the stamp. A reduced chi-square should strictly be normalised by $N_{pixels} - n - 1$, where n is the number of parameters in the kernel model, but since $N_{pixels} >> n$, we can overlook this. $T_i$, in the denominator of the summand, is the squared error term. This represents the squared Poisson or photon noise, which is the error one would expect in the limit of a perfect kernel solution. For TIC, $T$ and $R$ in equation~(\ref{diaqual}) are interchanged. Note that whilst equation~(\ref{diaqual}) is a variance-weighted statistic, the least-squares minimisation in ISIS~(v2.2) is performed without variance-weighting. However, ISIS does calculate variance-weighted $\chi^2_{red}$ values, which we use for our quality metric.

For each difference image, ISIS returns the mean, \diamean{}, and the standard deviation, \diascat{}, of the reduced chi-squared across all stamps within the image. We run ISIS using $10\times10$ stamp regions per image, with each stamp covering $31\times31$ pixels. The \diamean{} and \diascat{} statistics are used as a direct measure of the difference image quality. \diamean{} is an indicator of the global accuracy of the kernel solution averaged over the image. \diascat{} represents the variation in subtraction quality over different sub-regions of the image, and can be taken as the error on \diamean{} (i.e. the standard deviation of stamp $\chi^2_{red}$ values around \diamean{} for a given image). For a good difference image \diamean{} should be small (of order unity) and \diascat{} should be much smaller than \diamean{}.

To facilitate our further studies into what makes a good reference image, we will use \diamean{} and, initially, its error \diascat{} as a measure of difference imaging performance. For each survey dataset, we carry out DIA for every possible non-identical image pair under RIC ($R\otimes k - T$) and collate the \diamean{} and \diascat{} statistics. For a given reference, $R$, we then have the set of statistics which correspond to all other targets, $T$, under RIC. We can easily obtain the set of statistics for a given reference under TIC ($T\otimes k - R$) from the statistics already obtained from RIC (from image pairs where $T$ is the reference image and $R$ is the target). Note that in TIC, $R$ and $T$ are interchanged in equation~\ref{diaqual}, so the variance (in the denominator) is estimated from the same unconvolved reference for every difference image in a single run. This is in contrast to RIC, where the variance is estimated from a different unconvolved target for each difference image. It is therefore not meaningful to directly compare the performance of RIC and TIC DIA, although it is still valid to look for trends in each.

\begin{figure*}
\includegraphics[clip,trim = 25 280 80 280,width=\textwidth]{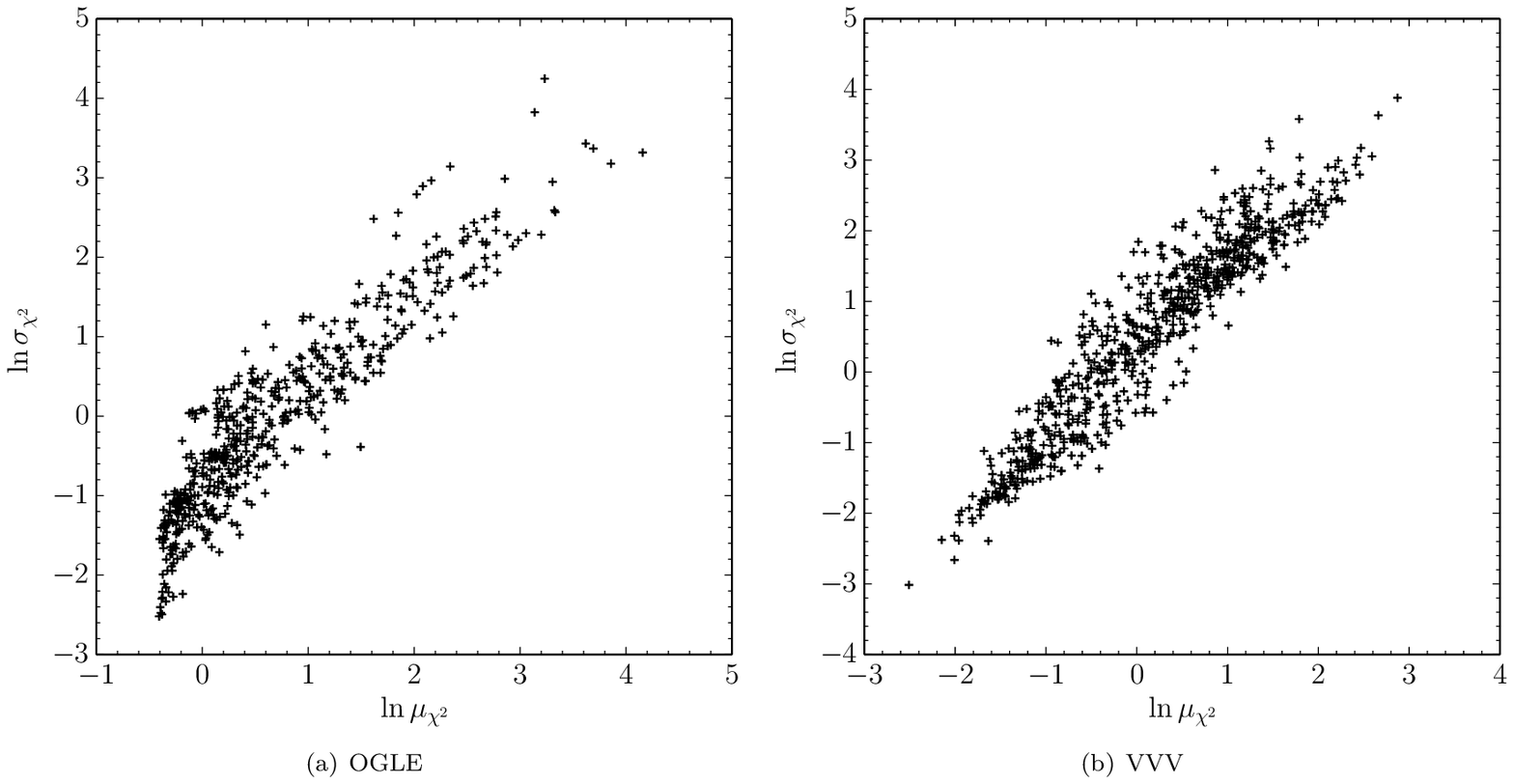}
\caption{$\ln{\diamean}$ against $\ln{\diascat}$ for (a) OGLE and (b) VVV from all possible image pairs under RIC ($R\otimes k - T$).}
\label{fig:meanscatterplots}
\end{figure*}

Figures \ref{fig:meanscatterplots}(a) and (b) show the values of \diamean{} and \diascat{}, in log-space, obtained for all image pairs in the OGLE and VVV test datasets, respectively. The two statistics show clear correlation for both surveys, as might be expected. The best statistics for OGLE are within the expected range for a reduced-$\chi^2$; \diamean{} has a minimum value near unity, and at least above 0.5 ($\ln{\diamean} > -0.69$). \diascat{} is comparatively small in this regime, down to 2 orders of magnitude less than \diamean{}. For VVV, \diamean{} reaches well below unity, despite having a correspondingly low \diascat{} in this regime. For a reduced-$\chi^2$, this would usually indicate an overestimated error term, but it may also be due to correlated noise between pixels during PSF-convolution.

\begin{figure*}
\includegraphics[clip,trim = 15 300 60 300,width=\textwidth]{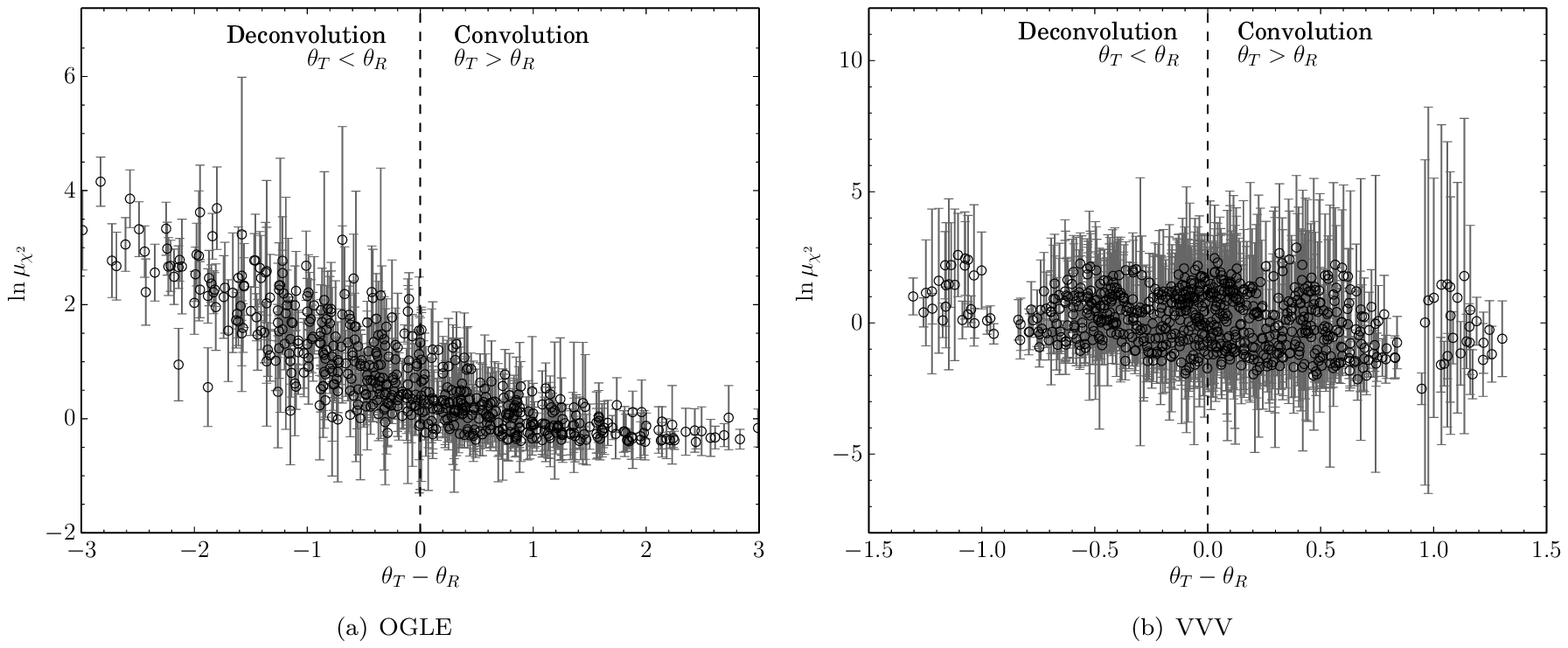}
\caption{Log mean DIA statistics, $\ln{\diamean}$, (black circles) for all image pairs under RIC ($R\otimes k - T$), plotted against the seeing difference, $\theta_T - \theta_R$, between the target, $T$, and the reference, $R$. Their proportional errors, $\diascat/\diamean$, are shown by the error-bars (in dark-grey, to highlight the distribution of the central points). The dashed line marks the separation between regimes where the reference, $R$, is convolved ($\theta_T > \theta_R$) or deconvolved ($\theta_T < \theta_R$) with the kernel, $k$, to match the target, $T$.}
\label{fig:seeingdiffplots}
\end{figure*}

In Figure \ref{fig:seeingdiffplots}(a) and (b), for OGLE and VVV, we plot $\ln{\diamean}$ against the difference in seeing, $\theta_T - \theta_R$, between target and reference for all image pairs under RIC ($R\otimes k - T$). The locus at $\theta_T - \theta_R = 0$ marks the transition where the reference, $R$, goes from being deconvolved to match the target ($\theta_T < \theta_R$) to where it is being convolved to match the target ($\theta_T > \theta_R$). For OGLE, where the data is well-sampled and has a broad range of seeing, there is a clear increase in $\ln{\diamean}$ going into the deconvolved regime, and a largely flat minimisation of $\ln{\diamean}$ in the convolved regime. The errors (proportional, since we are working with natural logs, i.e. $\diascat/\diamean$) are minimised at the tails of the distribution, where the seeing differences are greatest. VVV data is, by contrast, under-sampled, and occupies a smaller range of seeing. There is only a weak increase in $\ln{\diamean}$ going into the deconvolved regime. The largest errors are on a group of points at the largest and most positive end of the $\theta_T - \theta_R$ axis. This may be due to the VVV data being under-sampled. Any inaccuracies in the kernel solution caused by under-sampling the reference PSF may be amplified when convolving to a much broader seeing.

\begin{figure*}
\includegraphics[clip,trim = 15 300 50 320,width=\textwidth]{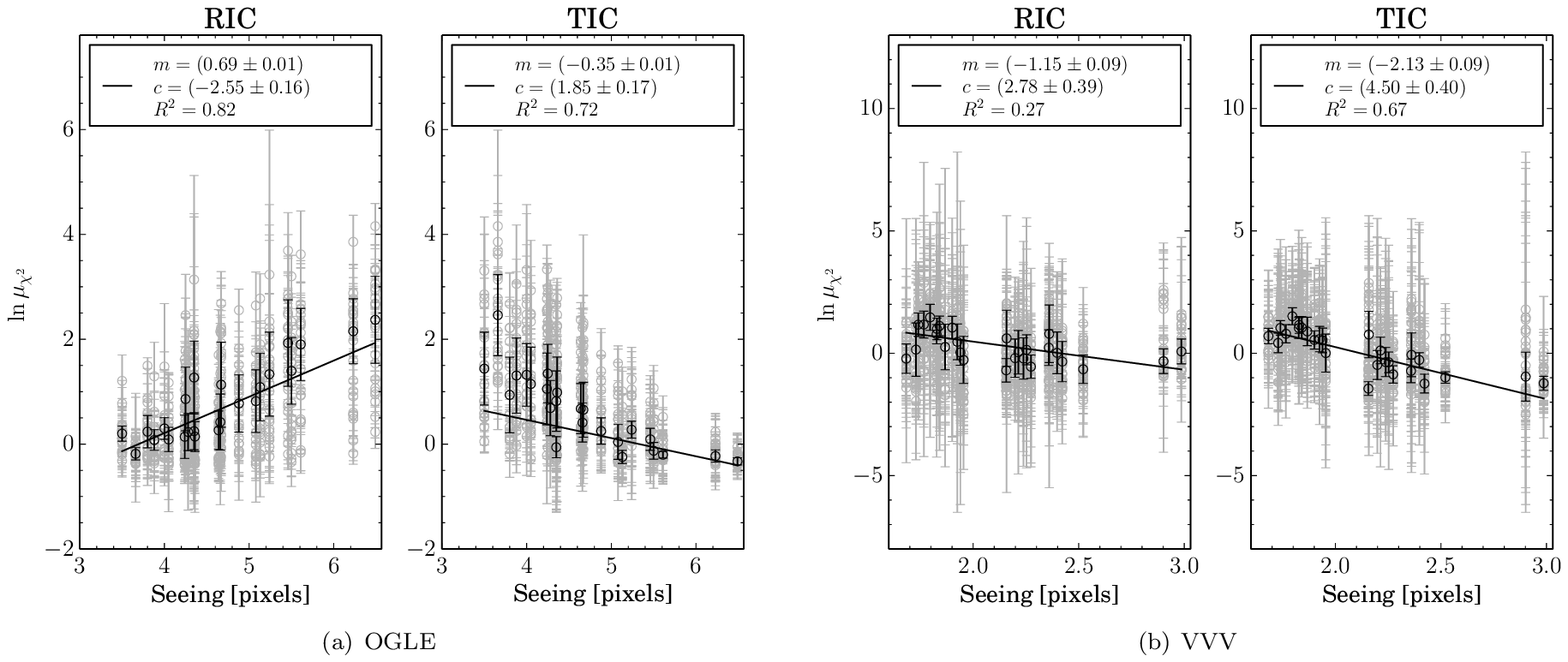}
\caption{Plots of $\ln{\diamean}$ DIA statistic against seeing for RIC and TIC for both survey datasets. In each case, the grey circles are the raw statistics for all image pairs, where the seeing axis corresponds to the seeing of the reference image (under RIC or TIC). The error-bars on these points represent the proportional error (on a log scale) from the scatter statistic, i.e. $\diascat/\diamean$. The black circles represent the \medlndiamean{} for all image pairs for each given reference image and its seeing. The error-bars on these points represent the median absolute deviation, \madlndiamean{}. The black line in each plot shows the result of a weighted least-squares regression over the \medlndiamean{} values. For each fit, the slope, $m$, and intercept, $c$, are shown in the legend along with their uncertainties and the coefficient-of-determination, $R^2$.}
\label{fig:seeingmeanplots}
\end{figure*}

In Figures~\ref{fig:seeingmeanplots}(a)~and~(b) for OGLE and VVV, reference image seeing is plotted against $\ln{\diamean}$ for both RIC and TIC. In each case, the raw image pair $\ln{\diamean}$ values are plotted as light grey circles, with their proportional errors, $\diascat/\diamean$. To get a measure of overall performance for a given reference image, we use the median value of $\ln{\diamean}$ for all image pairs using that reference. We do this in log-space because \diamean{} is bounded at zero. In Figures \ref{fig:seeingmeanplots}(a) and (b), these median values are plotted as black circles on a log-scale, with their error bars representing the median absolute deviation, \madlndiamean{}. We perform a weighted least-squares regression over the \medlndiamean{} values against reference image seeing to obtain an estimate of the degree of correlation. The slope, $m$, and intercept, $c$, of the regression line are given with their errors, along with the $R^2$ correlation coefficient, in the legend in each subplot in Figure~\ref{fig:seeingmeanplots}. In the OGLE data, we can see a strong correlation between seeing and \medlndiamean{} under RIC and a strong anti-correlation under TIC. This behaviour follows our expectations (\S\ref{sec:dia-tic}) for RIC, where a good seeing reference performs the best, and TIC, where a poor seeing reference image performs the best. In the VVV data, there is a poor \emph{anti}-correlation for RIC, with \medlndiamean{} being largely flat throughout the seeing scale. There is a stronger anti-correlation under TIC, with a slight improvement at large seeing, i.e. a decrease in \medlndiamean{}. The behaviour under RIC might be expected given that VVV seeing spans a narrower range, so that PSFs undergoing convolution do not change much in size, but also under-sampling may limit the quality of difference images obtained under RIC. It is notable that the largest-seeing reference image performs better under TIC than the smallest-seeing reference image under RIC, and that there is a stronger anti-correlation for TIC. Although there is some more significant scatter in the underlying image pairs at large seeing under TIC, the alternative method appears to perform better overall for under-sampled data.

\begin{figure*}
\includegraphics[clip,trim = 15 300 50 310,width=\textwidth]{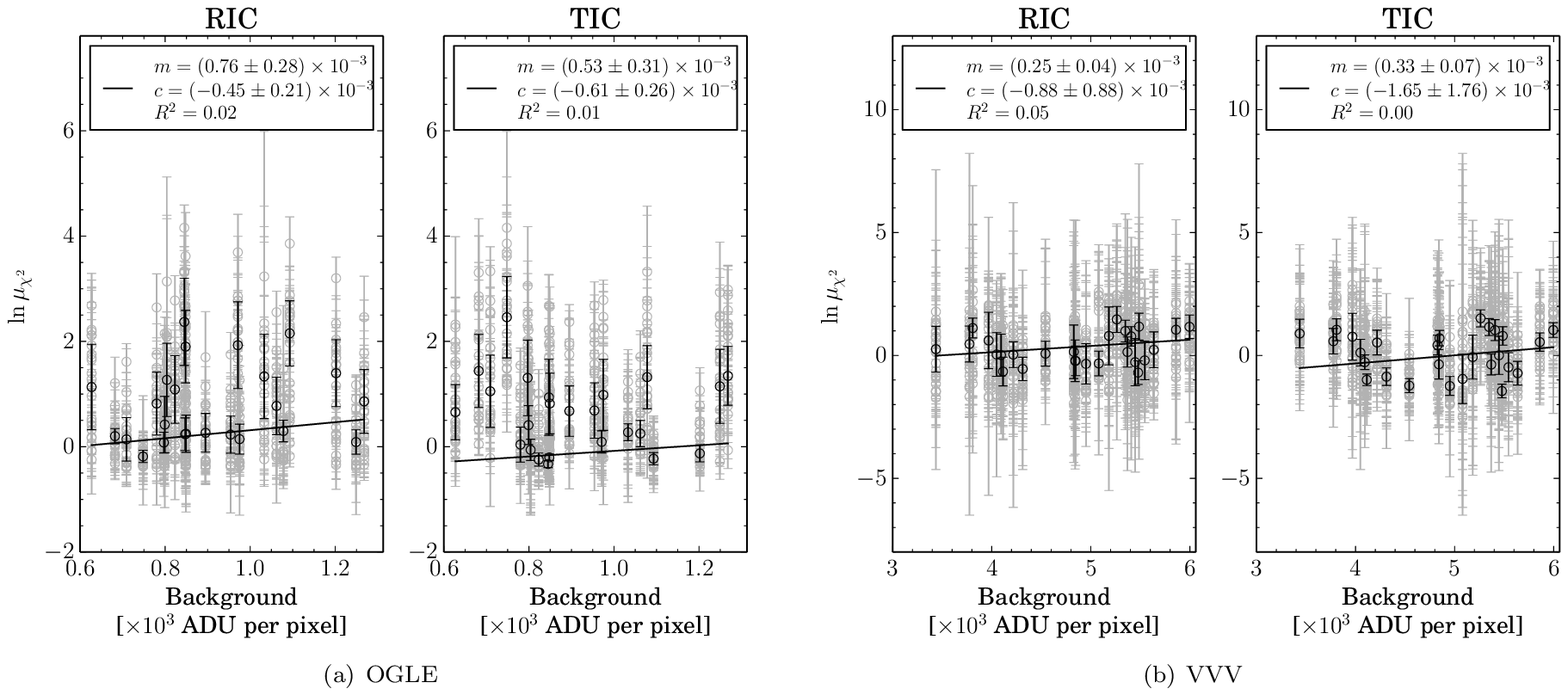}
\caption{As Figure \ref{fig:seeingmeanplots}, for reference-image background flux.}
\label{fig:bgmeanplots}
\end{figure*}

In Figures \ref{fig:bgmeanplots}(a) and (b), we plot $\ln{\diamean}$ against background flux for OGLE and VVV, under RIC and TIC. Linear regressions carried out for each survey in each convolution direction showed negligible correlation between \medlndiamean{} and background flux.

The results presented above demonstrate that under-sampled data perform differently under DIA to well-sampled data. They also show that TIC is a viable alternative DIA method, and is potentially optimal when dealing with under-sampled data.

In principle, through this analysis we have all the information needed to select our best reference image. For the well-sampled OGLE dataset, we might select the best-seeing reference image and use RIC DIA. For the under-sampled VVV data, we might choose a poorest-seeing reference image and use TIC DIA. However, to get to this conclusion, we have had to perform DIA on all possible reference--target image pairs in each dataset. For a larger dataset from a possibly ongoing survey, this approach may not be feasible. What we are aiming for is a more generalised reference image selection scheme which is a good predictor of which images would make the best reference image. The following sections, \S\ref{sec:ufom} and \S\ref{sec:sfom}, look at two possible approaches, using the DIA statistics computed in this section for calibration and evaluation.

\section{Approach 1: reference selection through a universal figure-of-merit}
\label{sec:ufom}

Perhaps the most ideal implementation of an automated reference selection scheme would be one which can be defined without having to be closely tailored to specifics of a particular survey, but instead reveals some universal scaling law dependent on only very basic and easily measurable image parameters. This would not only allow an easy way to select a reference image for a specific survey, it would also present a statistic which could be used to design future surveys which are optimised for difference imaging performance. In this section, we attempt to find such a scheme.

We propose fundamental criteria for the selection or construction of a reference image and provide a framework for quantitatively determining the relative importance of each criterion. These criteria are based on our ability to recover the PSF in the reference image under varying conditions of seeing and background. In principle, this is an open-ended problem as a full characterisation of a realistic PSF requires multiple free parameters (e.g. the standard OIS approach uses typically 6 parameters to define the shape of the local PSF kernel, with additional parameters to specify its variation over an image). However, we seek a method which is more broadly applicable and depends on only a few rapidly evaluated parameters. We therefore assume a simple 2D Gaussian model for the PSF.

\subsection{Reference image criteria} \label{refcrit}

A bare minimum set of parameters needed to define a PSF is a 2D position (centroid), width and amplitude. A circular Gaussian PSF is an example which would be completely characterised by this simple parameter set. We might expect the best reference image to be that for which the minimal PSF-parameter set is measurable to the best accuracy/precision out of all images in a given observing sequence. This would involve some overall evaluation which may involve some trade off between accuracies for individual parameters. The ability to measure these parameters depends in turn on basic characteristics of the image such as pixelisation (the seeing size relative to pixel scale), the background level and the source flux. For any real detector there will be a large number of additional factors which, in practise, limit the ability to characterise even simple Gaussian PSFs. However, we pursue a minimalist approach to reference image selection by considering our ability to characterise a simple 2D Gaussian PSF as a function of seeing \textit{in pixels}, background flux and, at least in general terms, source flux.

We set up large-scale simulations with our simplistic PSF model, spanning the parameter space of seeing and background, with a fixed PSF flux which represents the typical bright stars that OIS would use to determine the DIA convolution kernels. From these simulations, we construct maps which quantify the accuracy and precision with which we can measure the PSF as a function of seeing and background. We then use the locations of real images on these maps, along with their DIA statistics (see \S\ref{diaperf}), in order to calculate a Universal Figure-of-Merit (UFoM) which can be used to rank images by their expected success as a reference image. A good correlation between the UFoM and the DIA statistics would indicate that a reliable reference image selection algorithm can be based on this simple UFoM, which depends on a handful of basic, easily-measurable image parameters. If there is only a poor correlation, or no correlation, we can assume either that there is no such UFoM, the effect is too weak to be of practical use. It may also be the case that a successful UFoM would need to be rather more complex and/or involve more survey-dependent parameters than explored here.

\subsection{Universal Figure-of-Merit definition}
\label{sec:fomdef}

The metric space of seeing, background level (and representative source flux) is convenient in that these parameters can either be readily extracted from image meta-data (e.g. the headers of FITS-format images) or can be easily estimated from the image data themselves. For example, the background flux per pixel can be estimated as the mode of all pixel values, a reasonable estimator even in crowded fields. Since DIA routines such as OIS usually use the brightest unsaturated stars to compute the kernel transformation, our source flux should refer to the typical flux of ``bright stars'' on an image which could be approximated through simple flux summing (i.e. crude aperture photometry) of one or more unsaturated stars in an image.

It is important to stress that our search for a simple algorithm for good reference image selection should not be critically dependent on sophisticated and precise prescriptions for how quantities such as background level or bright star flux are measured. Otherwise it is unlikely that our method would be universal across different surveys or, worse, it may entail a computation which is as time consuming as manual image selection.

Once we have characterised our images, we want to derive for each image a UFoM score for reference image suitability. This UFoM could be calculated from a set of any number of \textit{heuristics}, $\left \{ H_i \right \}$, which quantify the accuracy with which we can characterise a reference image PSF. For our purposes the heuristics we choose will characterise the accuracy with which we can determine the PSF position, width and amplitude. Their precise form will be defined in \S\ref{sec:heuristics}.

We assume a general power-law form for the Universal FoM (UFoM) of:
\begin{equation}
UFoM \propto \prod_{i} H_i^{N_i} ,
\label{eq:ufom}
\end{equation}
where $N_i$ is a power ascribed to each heuristic $H_i$. In log space:
\begin{equation}
\ln(UFoM) \propto \sum\limits_{i} N_i\ln{\left ( H_i \right )}.
\label{eq:fom_log}
\end{equation}

With our UFoM, we wish to score potential reference images such that reference images with a higher score can be expected to produce better difference images. Since we will later calibrate it against real DIA \diamean{} statistics, it is useful to adopt the relation:
\begin{equation}
UFoM \propto \frac{1}{\diamean}.
\end{equation}

Therefore, where we believe that an increase in the value of a particular heuristic should improve reference image suitability and anti-correlate with \diamean{}, we assign it a positive power in equations \ref{eq:ufom} and \ref{eq:fom_log} (i.e. $N_i \geq 0$). Where we expect a particular heuristic to decrease with improving reference image suitability, such that it correlates with \diamean{}, we assign it a negative power ($N_i \leq 0$).

\subsection{PSF simulation}
\label{sec:psfsim}

For our PSF simulations we use a 2D Gaussian, sampled on a pixel grid of odd width, to reflect the form of the pixel stamp typically used in DIA to model the kernel (as described in \S\ref{sec:dia}). While a simple 2D Gaussian is not a perfect representation of a typical astronomical PSF (a \citealt{moffat1969} profile may be more realistic), it is the recoverability of the basic centroid, width and amplitude of a PSF that are of primary interest to us in searching for a simple reference image selection algorithm.

We construct a model PSF on a pixel grid such that the peak is in the central pixel. A random sub-pixel dither, ($x_{0,true}$, $y_{0,true}$), is applied to the position of this peak. The true width for the Gaussian PSF scales with input seeing, \seeing{}, as
\begin{equation}
w_{true} = \frac{\seeing}{2\sqrt{2\ln{2}}}
\end{equation}
and the true PSF amplitude is given by the input total stellar flux, $F_*$:
\begin{equation}
h=\frac{F_{*}}{2\pi w_{true}^2}.
\end{equation}
The expected pixel flux $E\left[F_i\right]$ in each pixel, $i$, on the grid is then:
\begin{equation}
E\left[F_i\right]=h\exp{\left[ - \frac{(x_i-x_{0,true})^2+(y_i-y_{0,true})^2}{2w_{true}^2} \right]} + F_{bg,true}~,
\end{equation}
where $F_{bg,true}$ is the input background flux per pixel and ($x_i$, $y_i$) are the coordinates at the centre of each pixel $i$. For input seeing $\seeing<3$~pixels, we sub-sample the model Gaussian from a grid of at least 3 times finer resolution to average over sub-pixel flux variation (the true, underlying PSF should be unaffected by pixelisation). Lastly, to account for photon noise, we replace the expected flux for each pixel, $E\left[F_i\right]$, with its Poisson realisation, $F_i$.

\subsection{Reference selection heuristics}
\label{sec:heuristics}

We have sampled our input Gaussian model over a finite pixel grid, added a background  level, applied a sub-pixel offset and included Poisson noise. We now wish to know what effect all of this has on our attempts to characterise the underlying PSF.

First, we define our measurable PSF parameters, namely the PSF centroid, ($x_0$, $y_0$), the Gaussian width, $w$, and the signal-to-noise, $s$ (which is related to the PSF amplitude expressed in units of background noise). For a single simulated PSF, the measured centroid is the flux-weighted centre:
\begin{equation}
x_0=\frac{\sum\limits_{i}F_{i}x_{i}}{\sum\limits_{i}F_{i}}~,~~~y_0=\frac{\sum\limits_{i}F_{i}y_{i}}{\sum\limits_{i}F_{i}}~.
\end{equation}

The Gaussian width is the combined RMS widths in $x$ and $y$:
\begin{equation}
w=\sqrt{\frac{\sum\limits_{i}\big((x_{i}-x_{0})^2+(y_{i}-y_{0})^2\big)F_{i}}{2\sum\limits_{i}F_{i}}}~. \label{eq:width}
\end{equation}

The signal-to-noise is the ratio of the flux contained in the PSF to the Poisson noise (from PSF and background) summed over the pixel grid:
\begin{equation}
s=\frac{\sum\limits_{i}{\left ( F_{i}-\fbg \right ) }}{\sqrt{\sum\limits_i{F_{i}}}} \label{eq:sn}
\end{equation}

The background flux, \fbg{}, is taken to be the median flux of the pixel grid, giving an estimated PSF flux of ${F_{PSF} = \sum_i (F_i - \fbg)}$. This provides a reasonable estimate of the input stellar flux provided the grid width is a few times larger than the seeing size.

We simulate PSFs over a grid of points spanning our metric space of seeing and background level. (We fix the value of the PSF flux at the median bright star flux of all images.) At each grid point we perform 1000 PSF simulations, each with random sub-pixel dithering, $(x_{0,true}$, $y_{0,true})$.

Our simulations must be compared to the real image data described in \S\ref{thedata} and therefore we also have to be mindful of saturation levels, especially for the VVV data. PSF simulations are dropped from our sample if the value of any of their pixels exceeds $90\%$ of the data saturation limit (to avoid the non-linearity regime; this reflects the limit we would impose on the bright stars used to fit the OIS convolution kernel). If less than $10\%$ of the original 1000-simulation sample remains, no calculation is made for that point in the metric space. Otherwise, from the remaining sample, we calculate the mean signal-to-noise, $\bar{s}$, for that point, and record the centroids and widths measured for the simulations. In practise, since we set the PSF flux from the median flux of unsaturated bright stars in all images, only significant outliers which are very poorly represented by this median value should fall in a saturated region of the metric space.

As high signal-to-noise in the PSF is clearly desirable in reference image selection, we choose the mean signal-to-noise, $\sn$, (across the sample at a given grid point) to be our first heuristic. For a second heuristic, we calculate a mean centroid accuracy, $\cacc$, from the measured and true centroids in each simulation, $k$:
\begin{equation}
\cacc=\frac{\sum\limits_k^N \sqrt{(x_{0,k}-x_{0,true,k})^2 + (y_{0,k}-y_{0,true,k})^2}}{N}
\end{equation}
This centroid accuracy measures how accurately we can locate the PSF on our pixel grid  given that we are limited by the pixel scale.

To take account of the effects of the pixel scale and seeing (and therefore spatial sampling) on determining the shape of the stellar PSF, we define two further heuristics based on the Gaussian width, which for a circular Gaussian contains all of the information about its shape. The third heuristic, width ``accuracy'', $\wacc$, is the average absolute deviation of the measured width, $w_k$, for a simulation, $k$, from the true width, $w_{true}$. This is normalised by the true width to obtain the relative accuracy:
\begin{equation}
\wacc=\frac{\sum\limits_k^N \lvert w_k-w_{true}\rvert}{N w_{true}}~.
\end{equation}

The fourth heuristic is the width ``precision'', $\wpre$, the standard deviation in the width measurements across the simulations:
\begin{equation}
\wpre = \sqrt{ \frac {\sum\limits_k^N \left(w_k - \bar{w} \right)^2}{N} }
\end{equation}

The width accuracy, $\wacc$, is dependent on the true width. The width precision, $\wpre$, is sensitive to the finite size of our pixel grid and therefore dependent on the background. In the regime of a 0-flux PSF, equation (\ref{eq:width}) will measure only a characteristic ``width'' of the pixel grid. At larger background, we essentially measure this characteristic width with better precision, since we have more ``photons'' in the array.

To calibrate $\wacc$ and $\wpre$, we generate a set of simulations across the same metric space where the PSF flux is set to zero (only a flat background level is present on the image grid). The calibrated heuristics, for each point in the metric space, are obtained through:
\be\label{eq:wacccal}
\wacccal = \frac{\wacc}{\wacczero}
\ee
and
\be\label{eq:wprecal}
\wprecal = \frac{\wpre}{\wprezero},
\ee
where $\wacczero$ and $\wprezero$ are the heuristics derived from the zero-flux simulations. All further instances in this paper of the symbols $\wacc$ and $\wpre$ can be assumed to be calibrated.

We expect that the signal-to-noise, $\sn$, should anti-correlate with the DIA statistic, \diamean{}, and correlate with the UFoM, while all other heuristics described above (centroid accuracy,~$\cacc$, width accuracy,~$\wacc$, and width precision,~$\wpre$) should correlate with \diamean{} and anti-correlate with the UFoM. We therefore assume the following limits on their powers when we calibrate the UFoM in equations \ref{eq:ufom} and \ref{eq:fom_log}:
\begin{equation}
N_{\sn} \geq 0;\quad N_{\cacc} \leq 0;\quad N_{\wacc} \leq 0; \quad N_{\wpre} \leq 0.
\end{equation}

\subsection{Heuristic measurements from real image data}

\begin{figure*}
\includegraphics[trim = 0 180 50 200, clip, width=\textwidth]{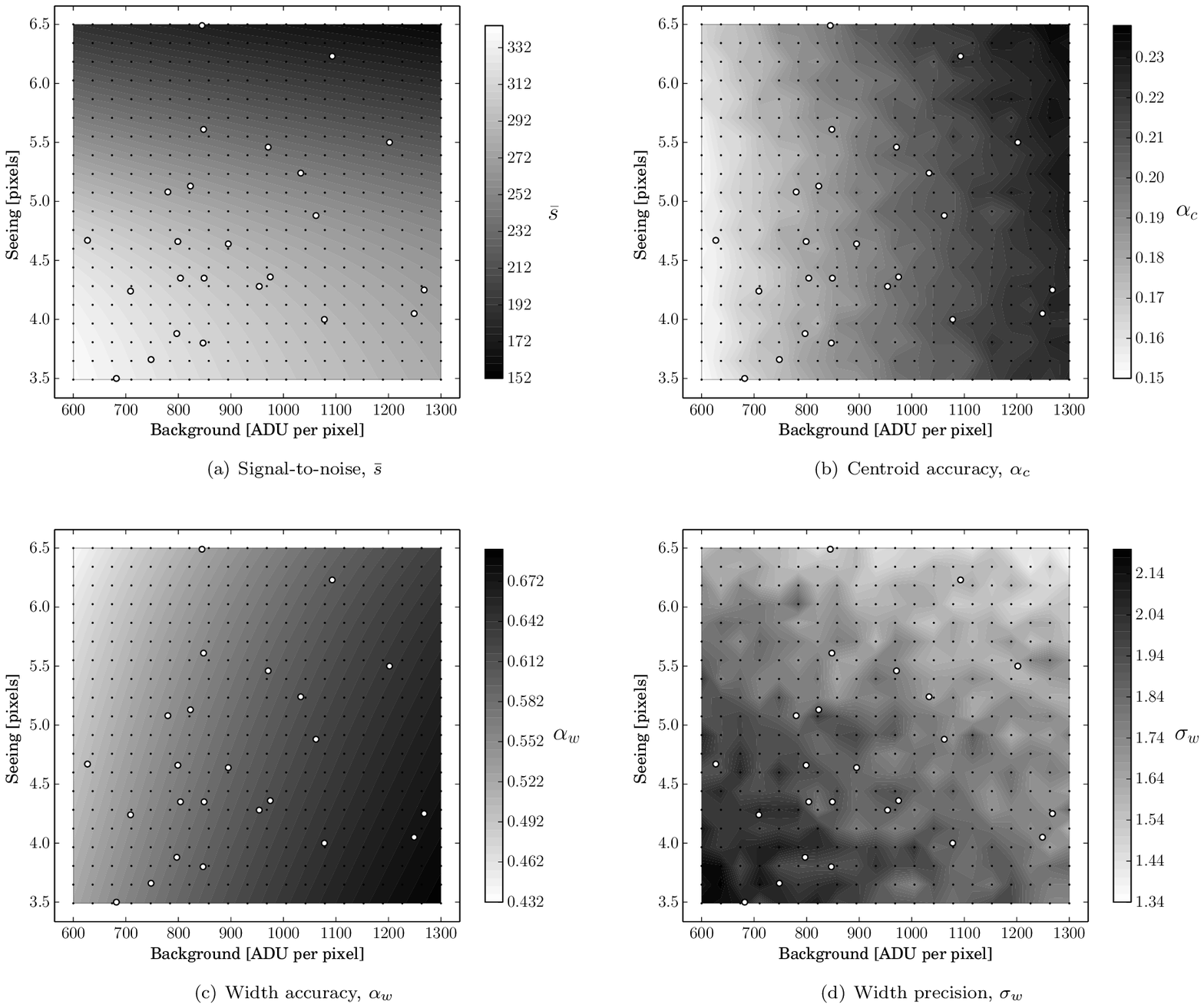}
\caption{Intensity maps of the four heuristics calculated over the metric space covered by the OGLE-III test data. Subfigures (a), (b), (c) and (d) are the signal-to-noise, $\sn$, centroid accuracy, $\cacc$, width accuracy, $\wacc$, and width precision, $\wpre$, respectively. Note that the greyscale is inverted for $\sn$, such that the four heuristics ``improve'' from black to white.  The grid of points at which simulations were carried out is shown by black dots. The positions of the images in the seeing-background space are shown by white circles.}
\label{fig:oglehplots}
\end{figure*}

\begin{figure*}
\includegraphics[clip,trim = 0 180 30 200,width=\textwidth]{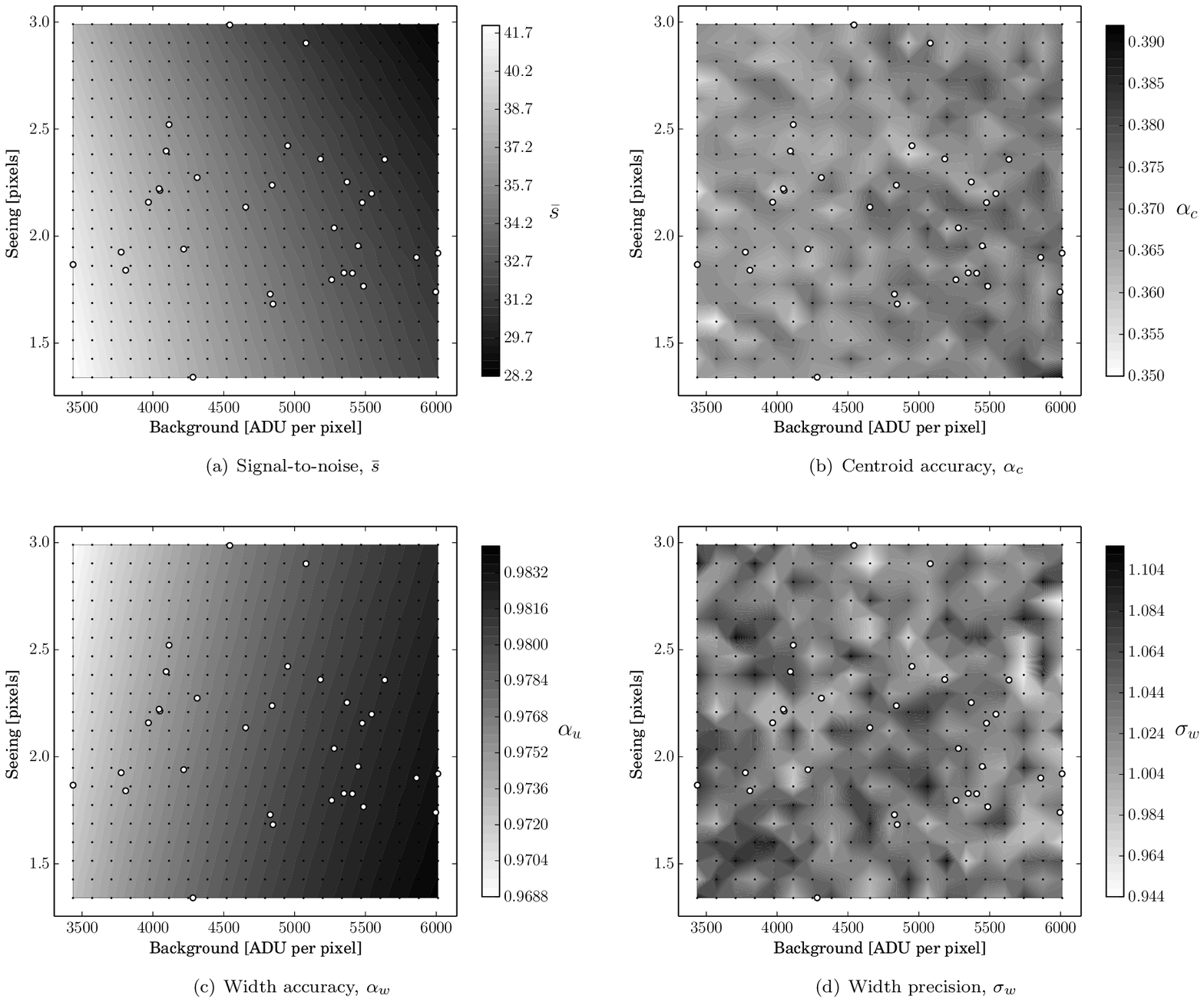}
\caption{As Figure \ref{fig:oglehplots}, for VVV test data.}
\label{fig:vvvhplots}
\end{figure*}

Figures~\ref{fig:oglehplots} and \ref{fig:vvvhplots} show, for OGLE and VVV respectively, maps of individual heuristic values of Gaussian PSF signal-to-noise ($\sn$), PSF centroid accuracy ($\cacc$), PSF width accuracy ($\wacc$) and PSF width precision ($\wpre$) over the plane of image seeing versus background level. In both cases the maps were constructed using a single bright star flux value $F_*$ which was set equal to the median flux value for each dataset (as reported in Table~\ref{tab:metrics}). The locations of the OGLE and VVV images in the metric space are shown in each figure.

For the heuristics derived from the OGLE simulations in Figure \ref{fig:oglehplots}, $\sn$ improves (increases) towards small seeing and low background, as might be expected. $\cacc$ improves (decreases) towards low background, as a result of lower noise levels. $\wacc$ improves (decreases) towards low background and large seeing, which may indicate that the greater sampling of the PSF, even in well-sampled data, can improve the (relative) accuracy of the PSF width determination. $\wpre$ shows some weak improvement (decrease) towards larger seeing and larger background. The effect of larger seeing might again be due to greater sampling of the PSF. The effect of larger background on the width precision heuristic is not clear to us; it could be that despite the calibration performed on this heuristic, the finite size of the pixel grid still has an effect, although we see no further way to mitigate this. Within the metric space occupied by the OGLE data, no PSF simulations contained saturated pixels. OGLE data has a high saturation limit and low typical background.

For VVV in Figure \ref{fig:vvvhplots}, both $\sn$ and $\wacc$ are broadly similar to those obtained for OGLE. $\sn$ improves (increases) towards lower background and small seeing whilst $\wacc$ improves (decreases) towards lower background and larger seeing. (Note that the metric spaces of the two surveys differ in dynamic range.) $\cacc$ is predominantly flat, but shows a hint of a rise in the corner towards small-seeing, large-background, where simulations are beginning to be dropped due to saturation. $\wpre$ is also fairly flat, with some degree of noise.

\subsection{Calculation of the UFoM}
\label{sec:linregfom}

We calculate the heuristic powers, $N_i$, in the UFoM (equations \ref{eq:ufom} and \ref{eq:fom_log}), via regression analysis in log-space. We standardise the \medlndiamean{} values to obtain the regressand, $y$ (we center on the mean and normalise by the standard deviation). We set its error, $\sigma_y$, by normalising \madlndiamean{} by the standard-deviation in \medlndiamean{}. We also standardise the heuristics in log-space to obtain the regressors, $X_i$:
\begin{equation} \label{eq:standardise}
X_i = \frac{\ln(H_i) - mean{\left(\ln(H_i)\right)} }{ std{ \left( \ln(H_i) \right) }}
\end{equation}
Standardising the quantities in this way normalises the differing scales of the heuristics and \medlndiamean{}, intrinsically to the survey. This should allow us to see if the UFoM is indeed universal across different surveys.

The regression problem is then to determine the \textit{absolute} gradients, $m_i$, which minimise the error function:
\begin{equation}
error = \frac{\left| y - \sum\limits_i n_i m_i X_i \right|}{\sigma_y},
\end{equation}
where $n_i$ represents the sign we ascribe to heuristic $i$ according to whether we believe it correlates ($n_i = 1$) or anti-correlates ($n_i = -1$) with $\ln{\diamean}$.

The heuristic powers, $N_i$, are obtained through:
\begin{equation}
N_i = - \frac{n_i m_i}{\sum\limits_i m_i }
\end{equation}
This normalisation of the dataset-specific values, $m_i$, is intended to better show that the \textit{relative} heuristic powers, $N_i$, are universal (if indeed they are). The additional minus sign causes the UFoM to be in anti-correlation with the DIA statistic; a larger UFoM score implies a smaller overall $\ln\diamean$.

\begin{figure*}
\includegraphics[clip,trim = 15 325 150 330,width=\textwidth]{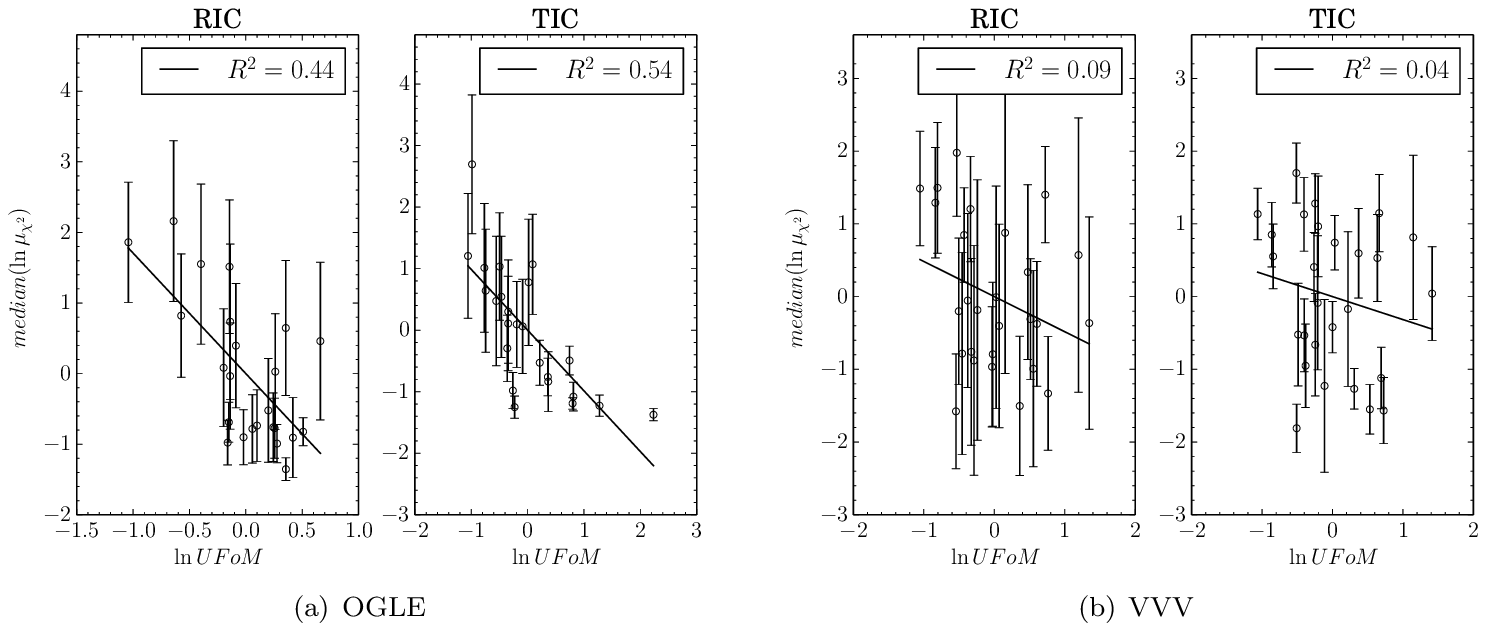}
\caption{Derived $\ln UFoM$ against \medlndiamean{} for OGLE and VVV under RIC and TIC DIA. The correlation coefficients for each derived UFoM are shown in the legends.}
\label{fig:ufomplots}
\end{figure*}

The derived heuristic powers, $N_i$, and their errors, are shown in Tables \ref{tab:ufom}(a) and (b) for OGLE and VVV, respectively. The corresponding UFoMs (in log-space) for RIC and TIC DIA are plotted in Figures \ref{fig:ufomplots}(a) and (b). The $\ln{UFoM}$ for OGLE shows a weak correlation ($R^2$) in either convolution direction, while for VVV, there is negligible correlation. Some of the derived heuristic powers for either RIC or TIC are consistent within the errors across the two datasets, but the errors are very large.

\begin{table*}		
\centering		
\caption{UFoM powers derived by linear regression, in log-space, of heuristics against \medlndiamean{} in each convolution direction. $R^2$ is the coefficient of determination in each case.}
\label{tab:ufom}
\begin{tabular}{lcccccc}
\multicolumn{6}{c}{(a) OGLE} \\
	&	$N_{\sn}$	&	$N_{\cacc}$	&	$N_{\wacc}$	&	$N_{\wpre}$	&	$R^2$ \\
\hline
RIC & $0.48 \pm 0.32$ & $-0.24 \pm 0.23$ & $-0.00 \pm 0.14$ & $-0.28 \pm 0.39$ & 0.44 \\
TIC & $0.00 \pm 1.38$ & $-0.00 \pm 2.48$ & $-0.23 \pm 1.89$ & $-0.77 \pm 0.40$ & 0.54 \\

\hline
\end{tabular}
\vspace{5mm}
\begin{tabular}{lcccccc}
\multicolumn{6}{c}{(b) VVV} \\
	&	$N_{\sn}$	&	$N_{\cacc}$	&	$N_{\wacc}$	&	$N_{\wpre}$	&	$R^2$ \\
\hline
RIC & $0.00 \pm 0.16$ & $-0.18 \pm 0.16$ & $-0.49 \pm 0.23$ & $-0.33 \pm 0.16$ & 0.09 \\
TIC & $0.00 \pm 0.25$ & $-0.26 \pm 0.21$ & $-0.40 \pm 0.32$ & $-0.34 \pm 0.15$ & 0.04 \\

\hline
\end{tabular}
\end{table*}

\section{Approach 2: reference selection through a survey-dependent figure-of-merit}
\label{sec:sfom}

An alternative approach to the UFoM described in the previous section, is simply to derive a survey-specific figure-of-merit (SFoM) directly from basic image properties. Such a figure-of-merit cannot be meaningfully compared between different surveys as it provides only an internal comparison of relative image performance. Nonetheless a reliable SFoM, which can be quickly evaluated from a small subset of survey data, is still a worthwhile goal for large-scale surveys in which manual selection of a reference image may be very time consuming.

The heuristics for this SFoM are taken to be simply the seeing, $\theta$, and background flux level, \fbg{}, both of which are obtained from image meta-data or directly measured from the image data. Following equations \ref{eq:ufom} and \ref{eq:fom_log}, we assume a power-law SFoM of the form:
\begin{equation}
SFoM \propto \theta^{N_{\theta}}\fbg^{N_{\fbg}},
\label{eq:sfom}
\end{equation}
or, in log-space:
\begin{equation}
\ln{SFoM} \propto N_{\theta}\ln{\theta} + N_{\fbg} \ln{\fbg}
\label{eq:sfom_log}
\end{equation}

For RIC, we expect DIA performance to improve with decreasing seeing, while for TIC we expect the opposite (due to better sampling of the PSF). In either convolution direction, we expect performance to improve with decreasing background flux. We therefore impose the following limits on the powers in the SFoM:
\begin{equation}
\begin{array}{rcr}
N_{\fbg} \leq 0;~~
\begin{cases}
	N_\theta \leq 0 & \text{if RIC} \\
	N_\theta \geq 0 & \text{if TIC}
\end{cases}
\end{array}
\end{equation}

Typical stellar flux is not used in this approach, since the SFoM is survey-specific and it is assumed that all images are of the same field of stars with the same flux distribution.

We undertake a regression analysis similar to that in \S{\ref{sec:linregfom}}, with reference image $\theta$ and \fbg{} replacing the heuristics, against  standardised \medlndiamean{} and its proportional error, $\madlndiamean/std\left(\medlndiamean\right)$. As before, $\ln{\theta}$ and $\ln{\fbg}$ are standardised prior to the linear regression (see equation \ref{eq:standardise}) to obtain $X_{\theta}$ and $X_{\fbg}$. In this case, we determine the absolute gradients $m_\theta$ and $m_{\fbg}$ which minimise:
\begin{equation}
error = \frac{\left| y - \left( n m_\theta X_{\theta} + m_{\fbg} X_{\fbg} \right) \right|}{\sigma_y},
\end{equation}
where:
\begin{equation}
n = 
	\begin{cases}
		+1 & \text{if RIC} \\
		-1 & \text{if TIC}
	\end{cases}
\end{equation}

\begin{figure*}
\includegraphics[clip,trim = 15 325 150 330,width=\textwidth]{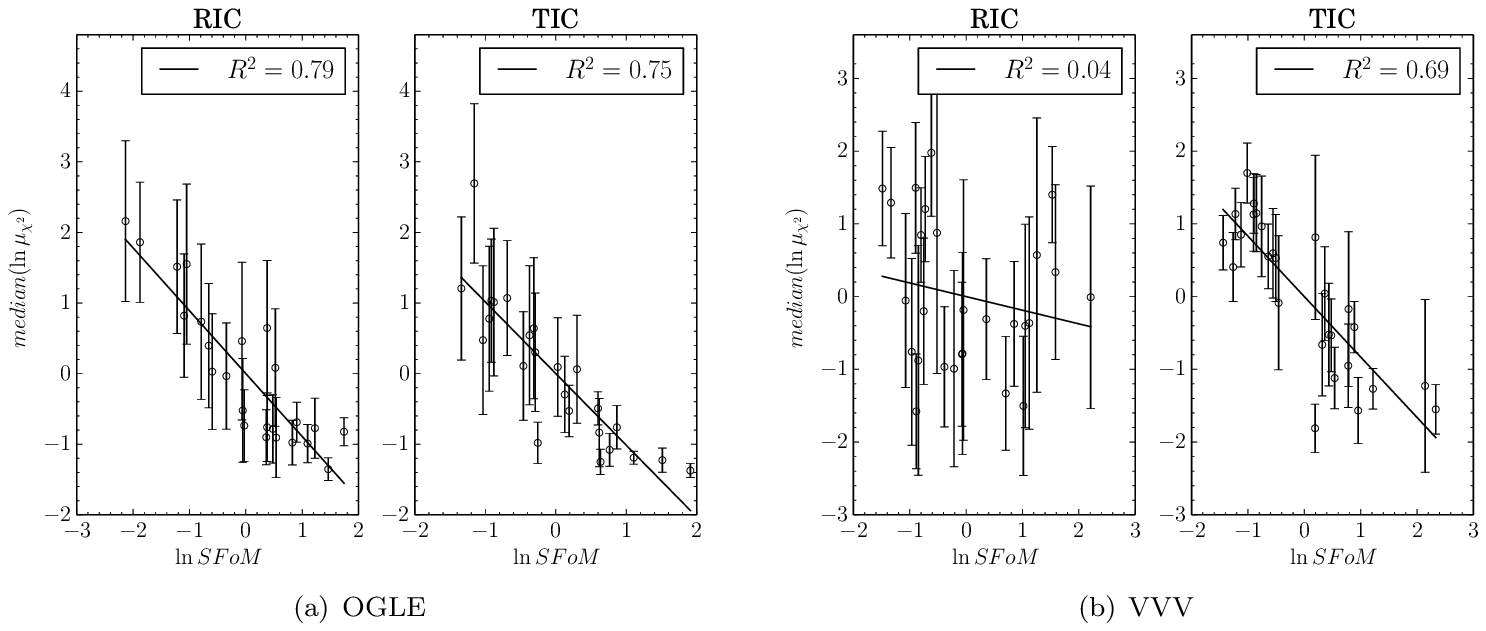}
\caption{Derived $\ln SFoM$ against \medlndiamean{} for OGLE and VVV under RIC and TIC DIA. The correlation coefficients for each derived SFoM are shown in the legends.}
\label{fig:sfomplots}
\end{figure*}

The derived powers for the SFoM power law are obtained from the fit parameters normalised by their sum in order to better see the relative contributions of $\theta$ and \fbg:
\begin{equation}
N_\theta = \frac{-n m_\theta}{m_\theta + m_{\fbg}};~~
N_{\fbg} = \frac{-m_{\fbg}}{m_\theta + m_{\fbg}}
\end{equation}
The values and errors for these powers are given in Tables \ref{tab:sfom}(a) and (b) for OGLE and VVV, respectively. The correlation of the SFoM for OGLE, under RIC and TIC, is fairly good, with seeing being dominant in both cases. Under TIC, it appears that background flux plays some small part in determining the SFoM. For VVV, the correlation is negligible under RIC, with the errors suggesting that the slope could lie anywhere. Under TIC, seeing is entirely dominant in the SFoM. The derived RIC and TIC SFoMs are plotted against \medlndiamean{} in Figures \ref{fig:sfomplots}(a) and (b) for OGLE and VVV, respectively.

\begin{table}		
\centering		
\caption{UFoM powers derived by linear regression, in log-space, of heuristics against \medlndiamean{} in each convolution direction. $R^2$ is the coefficient of determination in each case.}
\label{tab:sfom}
\begin{tabular}{lccc}
\multicolumn{4}{c}{(a) OGLE} \\
	&	$N_\theta$   &   $N_{F_{\rm bg}}$	&	$R^2$ \\
\hline
RIC & $-1.00 \pm 0.08$ & $0.00 \pm 0.16$ & 0.79 \\
TIC & $0.88 \pm 0.00$ & $-0.12 \pm 0.00$ & 0.75 \\

\hline
\end{tabular}
\vspace{5mm}
\begin{tabular}{lccc}
\multicolumn{4}{c}{(b) VVV} \\
	&	$N_\theta$   &   $N_{F_{\rm bg}}$	&	$R^2$ \\
\hline
RIC & $0.00 \pm 1.00$ & $-1.00 \pm 1.20$ & 0.04 \\
TIC & $1.00 \pm 0.12$ & $0.00 \pm 0.04$ & 0.69 \\

\hline
\end{tabular}
\end{table}

\section{Discussion}

We have examined the effect of seeing, background and pixel size on reference image selection in difference image analysis (DIA). We attempt to discover a universal figure-of-merit (UFoM) for selecting reference images for any survey. The heuristics we propose for the UFoM are based on the ability to characterise the reference image PSF under changing seeing and background conditions. Specifically, PSF signal-to-noise improves with decreasing seeing and background, while PSF width-accuracy improves with increasing seeing (due to greater sampling of the PSF) and decreasing background. The UFoM, however, is only weakly successful as a predictor of reference image performance in DIA for OGLE data, and is unsuccessful for VVV data. This may be due to the over-simplicity of our PSF model, although more complex PSF models would likely be too survey-specific and require many more model parameters.

We also investigated a survey-specific figure-of-merit (SFoM) based purely on seeing and background. As expected, seeing is the primary criterion in selecting a reference image and, for the well-sampled OGLE data, we find that standard difference imaging, which we term reference image convolution (RIC), obtains the best results. However, for our VVV dataset, which is usually spatially under-sampled, we find that reversing the standard procedure by convolving the target image to match a poorer seeing, but better-sampled reference, yields better quality difference images. We refer to this approach as target image convolution (TIC).

\citet{wozniak2008} advocates that the image point-spread function (PSF) should be sampled by at least 2.5~pixels per FWHM for good difference image results. Our tests are consistent with this where standard RIC DIA is performed. We show that TIC DIA provides a viable method in the event that most images fall below this limit. In either regime, the background level is an unimportant criterion, at least for the dynamic ranges of our test datasets, even in the case of near-infrared images from the VVV survey which have relatively high backgrounds compared to optical data. Therefore for both TIC and RIC methods, reference image performance is governed only by seeing, with best seeing reference images being preferred for RIC and worst seeing reference images for TIC. Of course it is possible that extremely poor seeing outside the range we considered would not have been favoured by our DIA quality statistics. We also point out that TIC may have other benefits, most notably allowing the difference images to be directly stacked onto the current reference to form a new high signal-to-noise reference frame, as all difference images are automatically PSF matched to their parent reference image.

Seeing remains the best criterion with which to predict reference image performance. In the well-sampled regime, this is the intuitive choice, although until now this had not been empirically shown to be the case. However, with standard methods (i.e. RIC DIA) it is not obvious from the outset what effect undersampling in the reference image might have on a series of difference images produced with it, where PSF-matching convolution kernels must be calculated with insufficient data. Given the potential impact of spatial-undersampling, we believe it is still worth considering the problem of reference image selection carefully before undertaking DIA for any survey.

\section{Summary}

We have compared the application of difference imaging analysis (DIA) to data where the PSF is spatially well-sampled (from the OGLE-III survey) and under-sampled (from the VVV survey). For under-sampled data, we propose an alternative method to standard DIA, in which the target images are convolved prior to subtraction from a poor-seeing reference image. We find this method is preferable when dealing with under-sampled data, compared to the standard method of reference-convolution with a good-seeing reference.

The selection of reference images for difference imaging analysis is an open-ended problem with no clear solution. We have made two attempts to find an algorithm for reference selection. In the first approach, we looked for a universal figure-of-merit which could be applied to any given survey. We based this on a set of heuristics, derived from simulations of a simple Gaussian PSF, which quantify the accuracy with which we can characterise the reference image PSF. We were not able to successfully derive such a figure-of-merit from our two test datasets (from the OGLE and VVV surveys). In our second approach, we attempted to find a survey-specific figure-of-merit based on simple image properties of seeing and background alone. This approach was more successful, finding good correlation with DIA performance for both datasets. It also points to the need for target-image convolution for under-sampled images.

\section*{Acknowledgments}
L.H. gratefully acknowledges the support of STFC and ESO through his studentships. We are very grateful to \L ukasz Wyrzykowski and the OGLE team for providing the OGLE-III data. We wish to thank the referee for their careful reading of this paper and several clarifying suggestions. This research made use of Astropy, a community-developed core Python package for Astronomy\footnote{See: http://www.astropy.org} \citep{astropy2013}.

\bibliographystyle{mn2e}

\begin{thebibliography}{}

\bibitem[\protect\citeauthoryear{Alard \& Lupton}{1998}]{alard1998} Alard C., Lupton R.~H., 1998, ApJ, 503, 325, arXiv:astro-ph/9712287

\bibitem[\protect\citeauthoryear{Alard}{2000}]{alard2000} Alard C., 2000, A\&AS, 144, 363

\bibitem[\protect\citeauthoryear{Astropy Collaboration}{2013}]{astropy2013} Astropy Collaboration, Robitaille, T.~P., Tollerud, E.~J., et al.\ 2013, A\&A, 558, A33, arXiv:1307.6212

\bibitem[\protect\citeauthoryear{Minniti et al.}{2010}]{minniti2010} Minniti D., et al., 2010, New Astron., 15, 433, arXiv:0912.1056

\bibitem[\protect\citeauthoryear{Moffat}{1969}]{moffat1969} Moffat A.~F.~J., 1969, A\&A, 3, 455

\bibitem[\protect\citeauthoryear{Saito et al.}{2012}]{vvvdr1} Saito R.~K., et al., 2012, A\&A, 537, A107, arXiv:1111.5511

\bibitem[\protect\citeauthoryear{Tomaney \& Crotts}{1996}]{tomaney1996} Tomaney A.~B., Crotts A.~P.~S., 1996, AJ, 112, 2872, arXiv:astro-ph/9610066

\bibitem[\protect\citeauthoryear{Udalski et al.}{2008}]{udalski2008} Udalski A., Szymanski M.~K., Soszynski I., Poleski R., 2008, Acta Astron., 58, 69, arXiv:0807.3884

\bibitem[\protect\citeauthoryear{Wozniak}{2008}]{wozniak2008} Wozniak P., 2008, in Kerins, E., Mao, S., Rattenbury, N. and Wyrzykowski, L., eds., Proceedings of the Manchester Microlensing Conference

\end{thebibliography}

\label{lastpage}

\end{document}